\documentclass[aps,prl,twocolumn,10pt,nofootinbib,superscriptaddress,preprintnumbers]{revtex4-2}

\usepackage[utf8]{inputenc}
\usepackage[T1]{fontenc}
\usepackage{amssymb,amsfonts,amsmath}
\usepackage{graphicx}
\usepackage{hyperref}
\usepackage{nicefrac}
\usepackage{xcolor}

\graphicspath{{plots/}}

\newcommand{\chiPT}{$\chi$PT}

\renewcommand{\l}{\left}
\renewcommand{\r}{\right}
\newcommand{\g}[1]{\gamma_{#1}} 
\newcommand{\trace}[1]{\mathrm{tr}\left[ #1 \right]} 
\newcommand{\tr}{\mathrm{tr}}
\newcommand{\<}{\langle}
\renewcommand{\>}{\rangle}

\newcommand{\gev}{\,\mathrm{GeV}}

\newcommand{\mev}{\,\mathrm{MeV}}
\newcommand{\fm}{\,\mathrm{fm}}

\newcommand{\SU}[1]{\mathrm{SU}\l(#1\r)}

\newcommand{\phys}{\mathrm{phys}}
\newcommand{\stat}[1]{\mathrm{stat}}
\newcommand{\sys}[1]{\mathrm{sys}}
\newcommand{\total}[1]{\mathrm{total}}
\newcommand{\staterr}[1]{(#1)_\mathrm{stat}}
\newcommand{\syserr}[1]{(#1)_\mathrm{sys}}
\newcommand{\totalerr}[1]{[#1]_\mathrm{total}}

\newcommand{\tins}{t_\mathrm{ins}}

\newcommand{\tsep}{t_\mathrm{sep}}
\newcommand{\tsepmin}{t_\mathrm{sep}^\mathrm{min}}
\newcommand{\tsepmax}{t_\mathrm{sep}^\mathrm{max}}

\newcommand{\rsqr}[3]{\langle r^2_{#1} \rangle_{#2}^{#3}}
\newcommand{\rsqrpisinglet}{\rsqr{S}{\pi}{0}}

\newcommand{\rsqrpi}{\rsqr{S}{\pi}{l}}
\newcommand{\FF}[4]{F_{#1}^{#2,#3}(#4)}

\newcommand{\FFpioctet}[1]{\FF{S}{\pi}{8}{#1}}
\newcommand{\FFpi}[1]{\FF{S}{\pi}{l}{#1}}

\newcommand{\mfp}{\mathfrak{p}}
\newcommand{\mfpi}{\mathfrak{p}_i}
\newcommand{\mfq}{\mathfrak{q}}
\newcommand{\mfpf}{\mathfrak{p}_f}
\newcommand{\vp}{\mathbf{p}}
\newcommand{\vpi}{\mathbf{p}_i}
\newcommand{\vq}{\mathbf{q}}
\newcommand{\vpf}{\mathbf{p}_f}
\newcommand{\vx}{\mathbf{x}}

\newcommand{\vxop}{\mathbf{x}_{op}}
\newcommand{\vxf}{\mathbf{x}_f}
\newcommand{\Ctwopt}[2]{C_{#1}^\mathrm{2pt}(#2)}

\newcommand{\Cthreeptf}[3]{C_{#1\mathcal{#2}^f#1}^\mathrm{3pt}(#3)}

\newcommand\PRISMA{\affiliation{PRISMA$^+$~Cluster~of~Excellence and Institut~f\"ur~Kernphysik, Johannes~Gutenberg-Universität~Mainz, 55099~Mainz, Germany}}
\begin{document}

\title{\textbf{Low-Energy Constants of Chiral Perturbation Theory from Pion
Scalar Form Factors in \texorpdfstring{$N_f=2+1$}{Nf=2+1}-Flavor Lattice QCD with Controlled Errors}}

\author{Georg von Hippel}   \email{hippel@uni-mainz.de} \PRISMA
\author{Konstantin~Ottnad}  \email{kottnad@uni-mainz.de}  \PRISMA

\preprint{MITP-25-025}

\date{\today}

\begin{abstract}
We determine the low-energy constants (LECs)
$f_0$, $L_4^r$ and $L_5^r$ of SU(3) Chiral Perturbation Theory (\chiPT{})
from a lattice QCD calculation of the scalar form factors of the pion
with fully controlled systematics.
Lattice results are computed on a large set of $N_f=2+1$ gauge ensembles
covering four lattice spacings $a\in[0.049,0.086]\mathrm{fm}$,
pion masses $M_\pi\in[130,350]\mathrm{MeV}$,
and various large physical volumes. By determining the notorious
quark-disconnected contributions with unprecedented precision and using a large
range of source-sink separations $\tsep\in[1.0,3.25]\fm$, we are able for the
first time to obtain the scalar radii from a $z$-expansion parameterization of
the form factors rather than a simple linear approximation at small momentum
transfer. The LECs are obtained from the physical extrapolation of the radii
using NLO SU(3) NLO \chiPT{} to parameterize the quark mass
dependence. Systematic uncertainties are estimated via model averages based
on the Akaike Information Criterion. Our determination
of $L_4^r$ is the first lattice determination to obtain a result not compatible
with zero.
\end{abstract}

\maketitle

\section{Introduction} \label{sec:introduction}

Quark-hadron duality implies that one can study low-energy hadron physics via
two different approaches, either on the phenomenological level of hadrons using
effective theories such as in Chiral Perturbation Theory (\chiPT), or
on the fundamental level of quarks and gluons using non-perturbative Quantum
Chromodynamics (QCD), in particular lattice QCD simulations.

Historically, \chiPT{} has been an indispensable tool for lattice QCD
practitioners, who required it to extrapolate results from lattice simulations
performed using unphysically heavy quark masses to the physical quark mass
point. Nowadays, however, the relationship between \chiPT{} and lattice QCD has
changed, since algorithmic improvements and the progress of computer technology
have enabled simulations directly at physical quark masses. Now it is lattice
QCD with its ability to simulate at unphysical values of the quark masses 
that can offer added value to \chiPT{} by extracting values for its low-energy
constants (LECs) from first principles.

Of particular interest in this context are quantities depending only
on a single or very fery LECs. An example of such quantities are the scalar radii
\begin{equation}
 \rsqr{S}{\pi}{f} = \l. - \frac{6}{F_S^{\pi,f}(0)} \frac{dF_S^{\pi,f}(Q^2)}{dQ^2} \r|_{Q^2\rightarrow 0}
\end{equation}
that parameterize the scalar form factors
\begin{equation}
 F^{\pi,f}_S(Q^2) = \l\langle \pi(\vpf) \r| \mathcal{S}^f \l| \pi(\vpi) \r\rangle
 \label{eq:scalar_FF}
\end{equation}
at low values of $Q^2=-q^2=-(p_f-p_i)^2$.
In the $N_f=2$ theory, the only isoscalar scalar density is the light one,
\begin{equation}
 \mathcal{S}^l = \bar{u}u + \bar{d}d \,,
 \label{eq:Sl}
\end{equation}
and in $\SU{2}$ \chiPT, the corresponding scalar radius depends only on
$\bar{\ell}_4$ via
\begin{equation}
 \rsqr{S}{\pi}{l} = \frac{1}{8\pi^2f_{\pi,\mathrm{phys}}^2}
 \left[-\frac{13}{2}+\bar{\ell}_4+\log\frac{M_{\pi,\mathrm{phys}}^2}{M_\pi^2}\right].
\end{equation}

With $N_f=2+1$ quark flavors, the scalar densities can be expressed in the
basis of the singlet and octet ones,
\begin{align}
 \mathcal{S}^0 &= \bar{u}u + \bar{d}d - 2\bar{s}s\label{eq:S8} \,, \\
 \mathcal{S}^8 &= \bar{u}u + \bar{d}d +  \bar{s}s\label{eq:S0} \,,
\end{align}
and in $\SU{3}$ \chiPT, the scalar radii are related by \cite{Gasser:1984ux}
\begin{align}
 \rsqr{S}{\pi}{0} &= \rsqr{S}{\pi}{8} + \delta r^2_S, \label{eq:rsqr0} \\
 \rsqr{S}{\pi}{l} &= \rsqr{S}{\pi}{8} + \frac{2}{3}\delta r^2_S, \label{eq:rsqrl}
\end{align}
where the octet radius depends only on $L_5^r$, while the singlet and light
radii depend on both $L_5^r$ and $L_4^r$.

The LECs $\bar{\ell}_4$, and $L_5^r$ and $L_4^r$ can therefore be accurately
determined from a high-precision determination of the scalar form factors.

In this letter, we present a high-statistics determination of the scalar form
factors with fully controlled systematics, including the extrapolation to the physical point.
The use of moving frames allows us to achieve far
better momentum resolution than previous studies,
and we determine the numerically challenging quark-disconnected with a
statistical precision that is more than an order of magnitude better than the
best preceding determination, resulting in the first lattice determination of
$L_4^r$ that is not compatible with zero.

\section{Setup} \label{sec:setup} 

The lattice calculation of the scalar form factors has been carried out on a
set of 17 gauge ensembles produced by the Coordinated Lattice Simulations (CLS)
consortium \cite{Bruno:2014jqa} using $N_f=2+1$ flavors of non-perturbatively
$\mathcal{O}(a)$-improved Wilson fermions \cite{Sheikholeslami:1985ij} and a
tree-level Symanzik-improved gauge action \cite{Luscher:1984xn}. The majority
of the ensembles use open boundary conditions in time to mitigate topological
freezing \cite{Luscher:2011kk,Luscher:2012av}, whereas some ensembles use
periodic boundary conditions. Due to the use of a twisted-mass regulator
\cite{Luscher:2012av} for the light quarks and a rational approximation for the
strange quark \cite{Clark:2006fx}, reweighting factors
\cite{Kuberski:2023zky,Bruno:2014jqa,Mohler:2020txx} have to be applied when
taking gauge averages. \par

Most of the ensembles lie on the chiral trajectory defined by $\tr[M] =
2m_l+m_s = \mathrm{const}$, but additional ensembles on a second chiral
trajectory defined by $m_s \approx m_s^\phys$ have been included in order to
provide a better handle on the separate $m_l$ and $m_s$ dependence
in the chiral extrapolations. \par

Scale setting is performed via the gradient flow scale $t_0$
\cite{Luscher:2010iy}, using the values for $t_0^\mathrm{sym}/a^2$ at the
symmetrical point from Ref.~\cite{Bruno:2016plf},
while defining the physical light and
strange quark masses using the $N_f=2+1$ FLAG world average 
\cite{FlavourLatticeAveragingGroupFLAG:2021npn}
\begin{equation}
 \sqrt{t_0^\phys}=0.14464(87)\fm \,,
 \label{eq:sqrt_t0phys}
\end{equation}

The matrix elements in Eq.~(\ref{eq:scalar_FF}) are evaluated in lattice QCD
by taking the ratio
\begin{align}
 \label{eq:ratio}
 &R^f(\mfpf, \mfq, \mfpi, \tsep, \tins) =
\frac{\<\Cthreeptf{P}{S}{\mfpf,\mfq,\mfpi, \tsep,
\tins}\>}{\<\Ctwopt{PP}{\mfpf, \tsep}\>} \\&\times
\sqrt{\frac{\<\Ctwopt{PP}{\mfpi, \tsep-\tins}\>
\<\Ctwopt{PP}{\mfpf^2, \tins}\> \<\Ctwopt{PP}{\mfpf^2,
\tsep}\>}{\<\Ctwopt{PP}{\mfpf^2, \tsep-\tins}\>
\<\Ctwopt{PP}{\mfpi^2, \tins}\> \<\Ctwopt{PP}{\mfpi^2,
\tsep}\>}} \,,\nonumber
\end{align}
of the two- and three-point functions
\begin{align}
 \Ctwopt{PP}{\vp, t} &= \sum_{\vxf} e^{i \vp\cdot\vxf} \l< P(\vxf, t) P^\dag(\mathbf{0}, 0) \r>_F \,, \label{eq:2pt} \\
 \Cthreeptf{P}{S}{\vpf, \vq, \tsep, \tins} &= \sum_{\vxf,\vxop} e^{i \vpf
\cdot \vxf} e^{i\vq \cdot \vxop} \times  \label{eq:3pt}\\&\l< P(\vxf, \tsep) \mathcal{S}^f(\vxop,
\tins) P^\dag(\mathbf{0}, 0) \r>_F \,, \nonumber
\end{align}
where $P(\vx, t) = \frac{1}{\sqrt{2}} \l[\bar{u}\g{5}u +
\bar{d}\g{5}d\r](\vx,t)$, and $\l(\mfpf,\mfq,\mfp_i\r)$ are equivalence classes
of lattice momenta $(\vpf,\vq,\vpi)$ over which the correlation functions have
been averaged.

Performing the Wick contractions for the fermionic expectation value $\<\cdot\>_F$ in the three-point function in Eq.~(\ref{eq:3pt})
yields both quark-connected and quark-disconnected diagrams.
The quark-connected piece is computed to high statistical precision using
a sequential propagator through the sink \cite{LHPC:2002xzk} and the truncated
solver method (TSM) \cite{Bali:2009hu,Blum:2012uh,Shintani:2014vja}.
The numerical evaluation of the quark-disconnected contribution requires
correlating two-point functions $\Ctwopt{PP}{\vpf, t}$ with scalar quark loops
\begin{equation}
 L_{\mathcal{S}^f}(\vq, t) = - \sum\limits_{\vx} e^{i \vq \cdot \vx}
\trace{D_f^{-1}(x,x)} \,,
\end{equation}
where $D_f$ is the Dirac operator for quark flavor $f=l,s$. These loops have
been calculated using the (OET+gHPE+HP)
prescription we introduced in Ref.~\cite{Ce:2022eix}
on the basis of the method of Ref.~\cite{Giusti:2019kff}, combining the one-end trick
(OET)~\cite{McNeile:2006bz} with the generalized hopping parameter expansion
(gHPE)~\cite{Gulpers:2013uca} and hierarchical probing
(HP)~\cite{Stathopoulos:2013aci}. \par

One final wrinkle concerns the estimation of the subtraction of the vacuum
expectation value (vev) of the scalar loop at zero momentum. With open boundary
conditions, the need to avoid the region close to the boundaries leads to
restrictions on the source positions that can be used at a given $\tsep$, which
in turn leads to an amplification of fluctuations in the loop at large $\tsep$,
where few sources contribute to the two-point function. To cancel these
fluctuations, we determine and subtract the vev on each timeslice separately,
leading to a large improvement in signal quality.

\section{Data Analysis}

For large time separations, the ratio~(\ref{eq:ratio}) tends directly to the
scalar form factor of the pion, $\lim_{\tsep\gg\tins\to\infty} R^f(\mfpf, \mfq,
\mfpi, \tsep, \tins) = F_S^{\pi,f}(Q^2)$.
To extract the form factor at finite $\tsep$, $\tins$, we use the summation
method~\cite{Maiani:1987by,Gusken:1989ad,Bulava:2011yz,Capitani:2012gj}
\begin{align}
 \sum_{\tins=\tau}^{\tsep-\tau} R^f(\mfpf, \mfq, \mfpi, \tsep, \tins) &=
C_\tau + \tsep F_S^{\pi,f}(Q^2) \label{eq:summation_method} \\&+
\mathcal{O}\left(e^{-\Delta\tsep}\right) \,, \nonumber
\end{align}
where $C_\tau$ is an irrelevant constant, and $\Delta$ is the energy gap between
the first excited state and the ground state. We perform linear fits
over different ranges $\tsep\in\left[\tsepmin,\tsepmax\right]$ of the summed
ratio, where $1.0\fm\lesssim\tsepmin\lesssim1.5\fm$ and
$2.5\fm\lesssim\tsepmax\lesssim3.0\fm$ on each ensemble, and
$\tau=\tsepmin/2$.

We parameterize the $Q^2$-dependence of the resulting form factors using the
$z$-expansion ansatz
\begin{equation}
  F_S^{\pi,f}(Q^2) = \sum_{n=0}^{N_z} a_n z^n\,, \quad
z=\frac{\sqrt{t_\mathrm{cut}+Q^2} -
\sqrt{\vphantom{Q^2}t_\mathrm{cut}-t_\mathrm{opt}}}{\sqrt{t_\mathrm{cut}+Q^2} +
\sqrt{\vphantom{Q^2}t_\mathrm{cut}-t_\mathrm{opt}}} \, ,
 \label{eq:z_expansion}
\end{equation}
in terms of which the radii are given by
\begin{equation}
 a_1 \sim \langle r_S^2\rangle_\pi^f = -\frac{6}{F_S^{\pi,f}(0)} \cdot
\left.\frac{dF_S^{\pi,f}(Q^2)}{dQ^2}\right|_{Q^2=0}\,.
 \label{eq:z_expansion_radii}
\end{equation}
In our fits, which we perform for five different cuts
$Q^2_\mathrm{max}\in\l[0.2,0.4\r]\gev^2$ on $Q^2\le Q^2_\mathrm{max}$,
we always use $N_z=1$, $t_\mathrm{cut}=4M_\pi^2$, and $t_\mathrm{opt} =
t_\mathrm{cut}(1-\sqrt{1+Q^2_\mathrm{max}/t_\mathrm{cut}})$ \cite{Lee:2015jqa}.

\begin{figure}[t]
 \centering
 \includegraphics[totalheight=0.22\textheight]{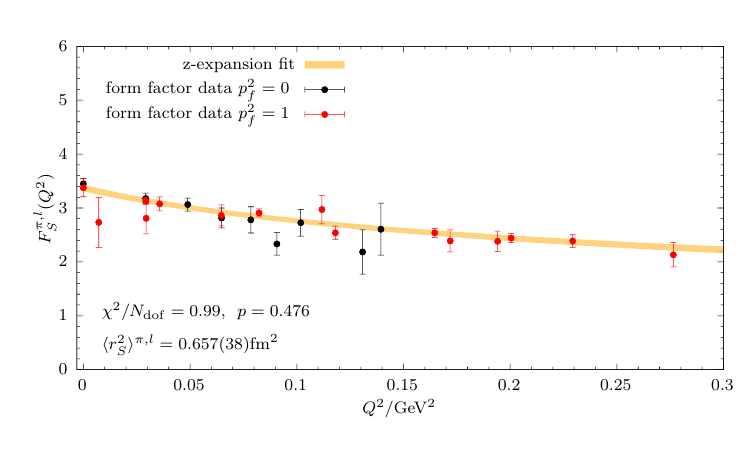}
 \includegraphics[totalheight=0.22\textheight]{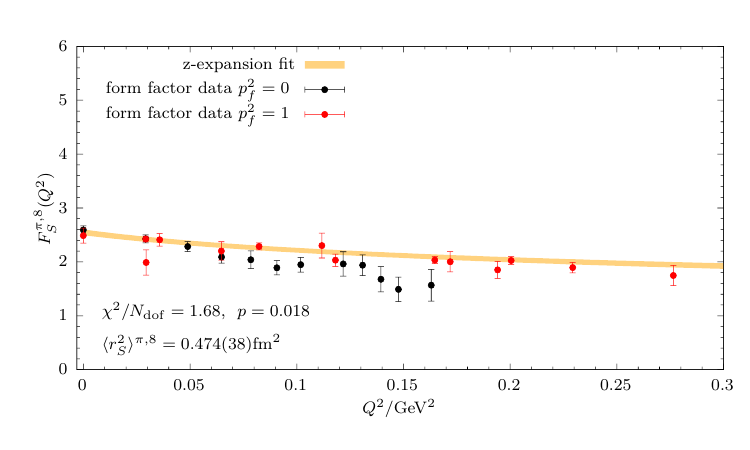}
 \caption{Form factor data (black for $\vpf^2=0$ and red for $\vpf^2=1$ frame)
          and $z$-expansion fit bands for $\FFpi{Q^2}$ (top)
          and $\FFpioctet{Q^2}$ (bottom) on the physical-mass ensemble E250.
          The results shown are for the data set with source-sink separations
          $\tsep\in[1.25,3.25]\fm$ in the summation method, and a fit range
          $Q^2\leq0.3\gev^2$ for the $z$-expansion.}
 \label{fig:FF_parametrization}
\end{figure}

Examples of the form factors obtained are shown in
Fig.~\ref{fig:FF_parametrization}. It can be clearly seen that the addition of
moving frames greatly improves the precision of the data and allows fits out to
much larger momentum transfers. 

To extrapolate the scalar radii to the physical point, we use fit ans\"atze
based on NLO $SU(3)$ $\chi$PT{}. To this
end, we rewrite the expressions for $\rsqr{S}{\pi}{8}$ and $\delta r^2_S$ in
terms of the leading-order quark-mass proxies $\xi_l=t_0 M_\pi^2$ and
$\xi_s=t_0(2M_K^2-M_\pi^2)$, expressing all dimensionful observables in units of
$t_0$ such that the scale-dependence $\mu$ is absorbed in the definitions of
$L_{4,5}^r$, implicitly setting $\sqrt{t_0} \mu = 1$ in the fits. We also
include a term $\sim a^2/t_0$ to account for discretization effects. To account
for finite-volume effects, we have also included the finite-volume corrections
\cite{Colangelo:2005gd}, but found that these tend to worsen the fit quality
while having no significant impact on the central values due to our already
rather large volumes.

While simultaneous fits for all three radii would in principle allow
simultaneous access to all LECs, we find that the very strong correlations
between the different radii render such fits problematic,
resulting in typically unacceptable fit qualities. We therefore opt to fit
suitable linear combinations that isolate $L_4^r$ and $L_5^r$.

We perform the entire analysis chain with different cuts on the range of
$\tsep$ used in the summation method, the range of $Q^2$ fitted in the
$z$-expansion fit, and the values of $a$, $M_\pi$ and $M_\pi L$ included in the
physical extrapolation. To arrive at our final best estimates including the
full statistical and systematic errors, we perform a model average
\cite{10.1177:0049124104268644,BMW:2014pzb}
based on the Akaike Information Criterion (AIC) \cite{Akaike:1974vps,Neil:2022joj}
by computing weights
\begin{equation}
w_{n,b} = \frac{e^{-B_{n,b}}}{N_B\sum_{k=1}^{N_M}e^{-B_{k,b}}}
\end{equation}
for model $n\in\{1,\ldots N_M\}$ on bootstrap resample $b\in\{1,\ldots,N_B\}$
with
\begin{equation}
B_{n,b} = \frac{1}{2}\chi^2_{n,b}+ N_{\mathrm{par},n}+N_{\mathrm{cut},n}-N_\mathrm{prio}
\end{equation}
where $\chi^2_{n,b}$ is the correlated $\chi^2$ for model $n$ on bootstrap
resample $b$, and $N_{\mathrm{par},n}$ and $N_{\mathrm{cut},n}$ are the number of
parameters in model $n$ and the number of data points cut for fitting with
model $n$, respectively \cite{Neil:2022joj}, while $N_\mathrm{prio}=1$ accounts for
the (uninformative) prior applied to $f_0$ to stabilize the fits.
Our empirical cumulative distribution function (CDF) is then
\begin{equation}
CDF(x) = \sum_{b=1}^{N_B}\sum_{n=1}^{N_M} w_{n,b}\Theta(x-x_{n,b})
\end{equation}
where $x_{n,b}$ is the value of $x$ obtained from model $n$ on bootstrap sample
$b$, and $\Theta$ is the Heaviside step function.

\begin{figure*}[t]
 \centering
 \includegraphics[totalheight=0.22\textheight]{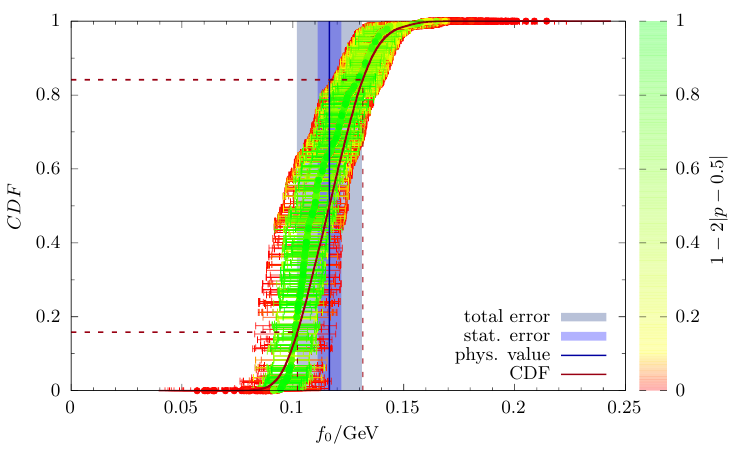}
 \includegraphics[totalheight=0.22\textheight]{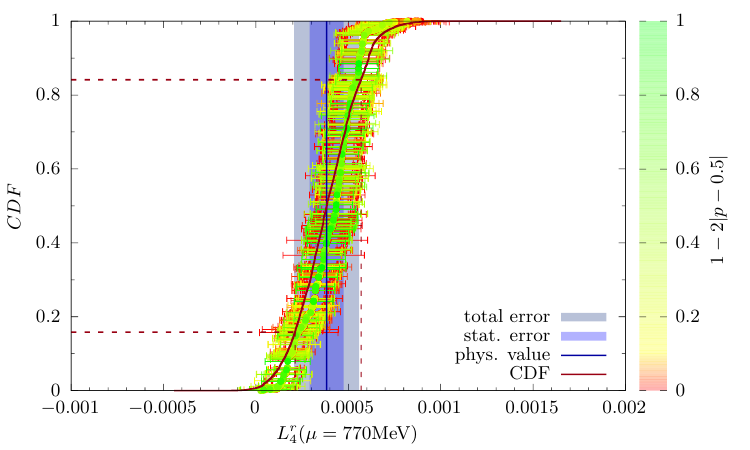} \\
 \includegraphics[totalheight=0.22\textheight]{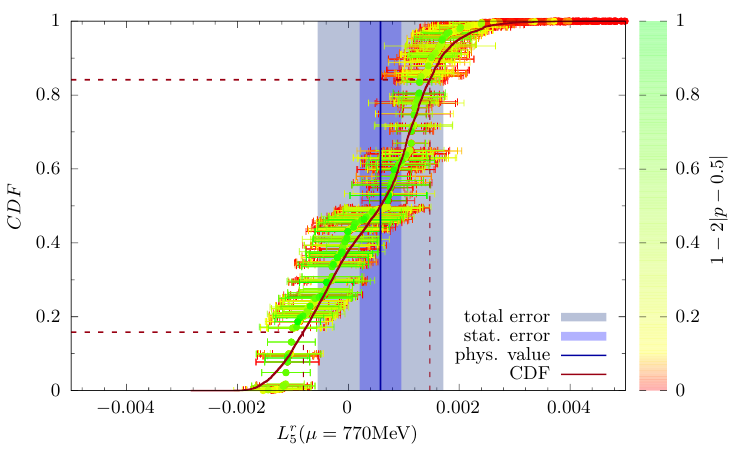}
 \includegraphics[totalheight=0.22\textheight]{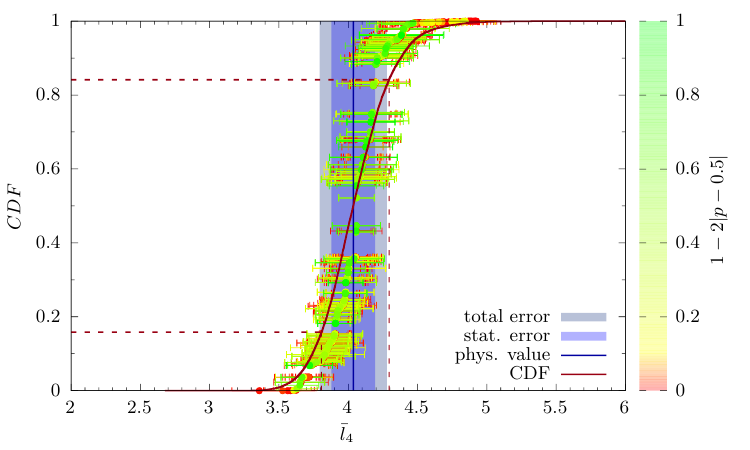}
  \caption{Cumulative distribution functions (CDFs) of the results for the
           $\SU{3}$ LECs $f_0$, $L_4^r(\mu)$ and $L_5^r(\mu)$ at a scale of
           $\mu=770\mev$ and the $\SU{2}$ LEC $\bar{\ell}_4$.
           Each data point represents the result and statistical
           error from an individual fit model and its color is determined by
           the corresponding $p$-value indicating the quality of the fit
           (which is different from the Akaike weight actually
           used to obtain the CDF). The solid vertical line indicates the final
           result from the model average, with the shaded bands giving its
           statistical and full errors, and the dashed lines corresponding to the
           $1\sigma$-quantiles of the CDF.}
 \label{fig:CDF_LECs}
\end{figure*}

Our final results for the LECs of $\SU{3}$ \chiPT{} are
\begin{align}
 f_0 &= 116.5\staterr{5.5}\syserr{13.7}\totalerr{14.8} \mev                           \,, \label{eq:f_0_phys} \\
 L_4^r(\mu) &= +0.38\staterr{09}\syserr{15}\totalerr{18} \times 10^{-3}       \,, \label{eq:L_4_phys} \\
 L_5^r(\mu) &= +0.58\staterr{0.38}\syserr{1.08}\totalerr{1.14} \times 10^{-3} \,, \label{eq:L_5_phys}
\end{align}
with the scale-dependent $L_i^r$ evaluated at a renormalization scale of
$\mu=770\mev$.
The corresponding CDFs are shown in the first three panels of
Fig.~\ref{fig:CDF_LECs}.

For the LO LEC $f_0$, our result is in excellent agreement with the FLAG
\cite{FlavourLatticeAveragingGroupFLAG:2021npn,MILC:2010hzw}
estimate $f_0 = 113.6(8.5)\mev$, albeit with
slightly larger errors. We note that the FLAG estimate derives
essentially from a single calculation of pion and kaon decay constants (not a
form factor calculation). We also agree almost perfectly with the recent
semiphenomenological estimate \cite{Lutz:2024ubv} $f_0 = 116.46(96)\mev$, 
while being in some tension with the result \cite{Liang:2021pql} of 
$f_0 = 82.3(14.1)\mev$ from the spectrum of the overlap Dirac operator.
For the NLO LEC $L_4^r$, our result is the first lattice result not to
be compatible with zero, to be compared with the FLAG estimate
\cite{FlavourLatticeAveragingGroupFLAG:2021npn,MILC:2010hzw}
$L_4^r(\mu)=-0.02(56) \times 10^{-3}$.
On the other hand, our form factor analysis is not able to obtain a
sufficiently precise value for $L_5^r$, where the FLAG estimate
\cite{FlavourLatticeAveragingGroupFLAG:2021npn,MILC:2010hzw} 
$L_5^r(\mu)=+0.95(41)\times 10^{-3}$ based
on an analysis of pion and kaon decay constants remains more accurate.

Fitting the $\SU{2}$ \chiPT{} formula to the results for the light scalar
radius, we obtain for the corresponding LEC
\begin{equation}
 \bar{\ell}_4 = 3.99\staterr{15}\syserr{17}\totalerr{23} \label{eq:l4bar_phys}
\end{equation}
in perfect agreement with, and about half the total error of, the FLAG
\cite{FlavourLatticeAveragingGroupFLAG:2021npn,MILC:2010hzw,Beane:2011zm,Borsanyi:2012zv,BMW:2013fzj,Boyle:2015exm}
estimate $\bar{\ell}_4 = 4.02(45)$. The corresponding CDF is shown in the last
panel of Fig.~\ref{fig:CDF_LECs}.

\section{Summary and discussion} \label{sec:summary}

We have obtained lattice results for the scalar form factors of the pion which
extend over a larger momentum range and have a much higher precision than any
previous study. As a result, we have been able to obtain the corresponding LECs of SU(2) and
SU(3) \chiPT{} with fully controlled errors, being able for the first time to
give an estimate for $L_4^r$ that is not compatible with zero.

We note that all existing determinations of $\SU{3}$ LECs as well as the
determinations of $\bar{\ell}_4$ using $N_f=2+1$ quark flavors are based on
decay constants, and thus involve two-point functions only, whereas we present the
first form factor calculation for $N_f=2+1$.

The LEC $L_4^r$, which we have been clearly able to distiguish from zero,
notably parameterizes a strange quark effect given by a purely
quark-disconnected contribution, that can be very cleanly determined from e.g.
$\rsqrpisinglet-\rsqrpi$ in our calculation, yielding a clear advantage for a
form factor calculation.

For $L_5^r$ on the other hand, our analysis does not have the same impact due
to, somewhat paradoxically, the high statistical precision to which the octet
radius determining it is computed: this leads to a large number of poor model
fits and a somewhat skewed CDF, in which the systematic error absolutely
dominates.

A more detailed description of our work and the form factors obtained is
forthcoming \cite{Ottnad:inpreparation}.

\section*{Acknowledgements}
The authors thank Stephan Dürr and Andreas Jüttner for useful comments and
discussions.
This research is supported by the Deutsche Forschungsgemeinschaft (DFG, German
Research Foundation) through project HI~2048/1-3 (project No.~399400745).
The authors gratefully acknowledge the Gauss Centre for Supercomputing e.V.
(www.gauss-centre.eu) for funding this project by providing computing time on
the GCS Supercomputer SuperMUC-NG at Leibniz Supercomputing Centre and on the
GCS Supercomputers JUQUEEN\cite{juqueen} and JUWELS\cite{JUWELS} at Jülich
Supercomputing Centre (JSC).
The authors gratefully acknowledge the computing time made available to them on
the high-performance computer Mogon-NHR at the NHR Centre NHR S\"ud-West.
This center is jointly supported by the Federal Ministry of Education and Research
and the state governments participating in the NHR (www.nhr-verein.de/unsere-partner).
Additional calculations have been performed on the HPC clusters Clover
at the Helmholtz-Institut Mainz and Mogon II and HIMster-2
at Johannes-Gutenberg Universit\"at Mainz.
The QDP++ library \cite{Edwards:2004sx} and the deflated SAP+GCR solver from the openQCD package \cite{openQCD} have been used in our simulation code.
We thank our colleagues in the CLS initiative for sharing gauge ensembles.
\bibliographystyle{h-physrev}
\bibliography{refs}

\begin{thebibliography}{10}

\bibitem{Gasser:1984ux}
J.~Gasser and H.~Leutwyler,
\newblock Nucl. Phys. B {\bfseries 250}, 517 (1985).

\bibitem{Bruno:2014jqa}
M.~Bruno {\em et~al.},
\newblock JHEP {\bfseries 02}, 043 (2015), 1411.3982.

\bibitem{Sheikholeslami:1985ij}
B.~Sheikholeslami and R.~Wohlert,
\newblock Nucl. Phys. B {\bfseries 259}, 572 (1985).

\bibitem{Luscher:1984xn}
M.~L\"uscher and P.~Weisz,
\newblock Commun. Math. Phys. {\bfseries 98}, 433 (1985),
\newblock [Erratum: Commun.Math.Phys. 98, 433 (1985)].

\bibitem{Luscher:2011kk}
M.~L\"uscher and S.~Schaefer,
\newblock JHEP {\bfseries 07}, 036 (2011), 1105.4749.

\bibitem{Luscher:2012av}
M.~L\"uscher and S.~Schaefer,
\newblock Comput. Phys. Commun. {\bfseries 184}, 519 (2013), 1206.2809.

\bibitem{Clark:2006fx}
M.~A. Clark and A.~D. Kennedy,
\newblock Phys. Rev. Lett. {\bfseries 98}, 051601 (2007), hep-lat/0608015.

\bibitem{Kuberski:2023zky}
S.~Kuberski,
\newblock Comput. Phys. Commun. {\bfseries 300}, 109173 (2024), 2306.02385.

\bibitem{Mohler:2020txx}
D.~Mohler and S.~Schaefer,
\newblock Phys. Rev. D {\bfseries 102}, 074506 (2020), 2003.13359.

\bibitem{Luscher:2010iy}
M.~L\"uscher,
\newblock JHEP {\bfseries 08}, 071 (2010), 1006.4518,
\newblock [Erratum: JHEP 03, 092 (2014)].

\bibitem{Bruno:2016plf}
M.~Bruno, T.~Korzec, and S.~Schaefer,
\newblock Phys. Rev. D {\bfseries 95}, 074504 (2017), 1608.08900.

\bibitem{FlavourLatticeAveragingGroupFLAG:2021npn}
Flavour Lattice Averaging Group (FLAG), Y.~Aoki {\em et~al.},
\newblock Eur. Phys. J. C {\bfseries 82}, 869 (2022), 2111.09849.

\bibitem{LHPC:2002xzk}
LHPC, TXL, D.~Dolgov {\em et~al.},
\newblock Phys. Rev. D {\bfseries 66}, 034506 (2002), hep-lat/0201021.

\bibitem{Bali:2009hu}
G.~S. Bali, S.~Collins, and A.~Schäfer,
\newblock Comput. Phys. Commun. {\bfseries 181}, 1570 (2010), 0910.3970.

\bibitem{Blum:2012uh}
T.~Blum, T.~Izubuchi, and E.~Shintani,
\newblock Phys. Rev. D {\bfseries 88}, 094503 (2013), 1208.4349.

\bibitem{Shintani:2014vja}
E.~Shintani {\em et~al.},
\newblock Phys. Rev. D {\bfseries 91}, 114511 (2015), 1402.0244.

\bibitem{Ce:2022eix}
M.~C\`e {\em et~al.},
\newblock JHEP {\bfseries 08}, 220 (2022), 2203.08676.

\bibitem{Giusti:2019kff}
L.~Giusti, T.~Harris, A.~Nada, and S.~Schaefer,
\newblock Eur. Phys. J. C {\bfseries 79}, 586 (2019), 1903.10447.

\bibitem{McNeile:2006bz}
UKQCD, C.~McNeile and C.~Michael,
\newblock Phys. Rev. D {\bfseries 73}, 074506 (2006), hep-lat/0603007.

\bibitem{Gulpers:2013uca}
V.~G\"ulpers, G.~von Hippel, and H.~Wittig,
\newblock Phys. Rev. D {\bfseries 89}, 094503 (2014), 1309.2104.

\bibitem{Stathopoulos:2013aci}
A.~Stathopoulos, J.~Laeuchli, and K.~Orginos,
\newblock SIAM J. Sci. Comput. {\bfseries 35}, S299 (2013), 1302.4018.

\bibitem{Maiani:1987by}
L.~Maiani, G.~Martinelli, M.~L. Paciello, and B.~Taglienti,
\newblock Nucl. Phys. B {\bfseries 293}, 420 (1987).

\bibitem{Gusken:1989ad}
S.~Güsken {\em et~al.},
\newblock Phys. Lett. B {\bfseries 227}, 266 (1989).

\bibitem{Bulava:2011yz}
J.~Bulava, M.~Donnellan, and R.~Sommer,
\newblock JHEP {\bfseries 01}, 140 (2012), 1108.3774.

\bibitem{Capitani:2012gj}
S.~Capitani {\em et~al.},
\newblock Phys. Rev. D {\bfseries 86}, 074502 (2012), 1205.0180.

\bibitem{Lee:2015jqa}
G.~Lee, J.~R. Arrington, and R.~J. Hill,
\newblock Phys. Rev. D {\bfseries 92}, 013013 (2015), 1505.01489.

\bibitem{Colangelo:2005gd}
G.~Colangelo, S.~Dürr, and C.~Haefeli,
\newblock Nucl. Phys. B {\bfseries 721}, 136 (2005), hep-lat/0503014.

\bibitem{10.1177:0049124104268644}
K.~P. Burnham and D.~R. Anderson,
\newblock Sociol. Methods \& Research {\bfseries 33}, 261 (2004).

\bibitem{BMW:2014pzb}
BMW, S.~Borsányi {\em et~al.},
\newblock Science {\bfseries 347}, 1452 (2015), 1406.4088.

\bibitem{Akaike:1974vps}
H.~Akaike,
\newblock IEEE Trans. Automatic Control {\bfseries 19}, 716 (1974).

\bibitem{Neil:2022joj}
E.~T. Neil and J.~W. Sitison,
\newblock Phys. Rev. D {\bfseries 109}, 014510 (2024), 2208.14983.

\bibitem{MILC:2010hzw}
MILC, A.~Bazavov {\em et~al.},
\newblock PoS {\bfseries LATTICE2010}, 074 (2010), 1012.0868.

\bibitem{Lutz:2024ubv}
M.~F.~M. Lutz, Y.~Heo, and R.~J. Hudspith,
\newblock Phys. Rev. D {\bfseries 110}, 094046 (2024), 2406.07442.

\bibitem{Liang:2021pql}
\ensuremath{\chi}QCD, J.~Liang {\em et~al.},
\newblock Phys. Rev. D {\bfseries 110}, 094513 (2024), 2102.05380.

\bibitem{Beane:2011zm}
S.~R. Beane {\em et~al.},
\newblock Phys. Rev. D {\bfseries 86}, 094509 (2012), 1108.1380.

\bibitem{Borsanyi:2012zv}
S.~Borsányi {\em et~al.},
\newblock Phys. Rev. D {\bfseries 88}, 014513 (2013), 1205.0788.

\bibitem{BMW:2013fzj}
BMW, S.~D\"urr {\em et~al.},
\newblock Phys. Rev. D {\bfseries 90}, 114504 (2014), 1310.3626.

\bibitem{Boyle:2015exm}
P.~A. Boyle {\em et~al.},
\newblock Phys. Rev. D {\bfseries 93}, 054502 (2016), 1511.01950.

\bibitem{Ottnad:inpreparation}
G.~M. von Hippel and K.~Ottnad,
\newblock in preparation.

\bibitem{juqueen}
{J\"{u}lich Supercomputing Centre},
\newblock Journal of large-scale research facilities {\bfseries 1} (2015).

\bibitem{JUWELS}
{J\"{u}lich Supercomputing Centre},
\newblock Journal of large-scale research facilities {\bfseries 7} (2021).

\bibitem{Edwards:2004sx}
SciDAC, LHPC, UKQCD, R.~G. Edwards and B.~Joo,
\newblock Nucl. Phys. B Proc. Suppl. {\bfseries 140}, 832 (2005),
  hep-lat/0409003.

\bibitem{openQCD}
M.~Lüscher {\em et~al.},
\newblock openqcd,
\newblock http://luscher.web.cern.ch/luscher/openQCD/.

\end{thebibliography}

\clearpage

\end{document}